\documentclass[usenatbib]{mn2e}
\usepackage{amsmath}
\usepackage{amssymb}
\usepackage{graphicx}
\usepackage{aas_macros}

\begin{document}

\author[Duffell \& MacFadyen]{Paul C. Duffell and Andrew I. MacFadyen\\
    Center for Cosmology and Particle Physics, Physics Department, New York University, New York, NY 10003, USA}
\title{High-Frequency Voronoi Noise Reduced by Smoothed Mesh Motion}
\maketitle


\begin{abstract}

We describe a technique for improving the performance of hydrodynamics codes which employ a moving Voronoi mesh.  Currently, such codes are susceptible to high-frequency noise produced by rapid adjustments in the grid topology on the smallest scales.  The treatment for this grid noise is simple; instead of moving the mesh-generating marker points with the local fluid velocity, this velocity field is smoothed on small scales, so that neighboring marker points generally have similar velocities.  We demonstrate significant improvement gained by this adjustment in several code tests relevant to the physics which moving-mesh codes are designed to capture.

\end{abstract}


\section{Introduction}
\label{sec:intro}

The fields of astrophysics and cosmology have benefited from the recent development of a numerical technique for effectively Lagrangian integration of the equations of hydrodynamics.  This moving-mesh technique, first proposed by \cite{1985LNP...238...87B} and recently implemented in the cosmology code AREPO \citep{2010MNRAS.401..791S} and later  for relativistic astrophysics in the TESS code \citep{2011ApJS..197...15D}, has had considerable success in improving upon results from Lagrangian Smoothed Particle Hydrodynamics (SPH) methods and Eulerian Adaptive Mesh Refinement (AMR) techniques.    Both TESS and AREPO are three-dimensional and parallel, and have implemented solvers for the magnetohydrodynamic and viscous hydro equations.  A related finite-element approach has also recently been developed which employs the same moving Voronoi mesh \citep{2014MNRAS.437..397M}.  The numerical method has also been adapted for use with alternate, non-Voronoi meshes tailored to disks and jets \citep{2012ApJ...755....7D, 2013ApJ...775...87D}, as the general idea does not depend sensitively on the shape of the zones.

In general, a moving mesh can be advantageous for capturing highly supersonic flows, or situations which require precise preservation of contact discontinuities.  One of the most highly-prized advantages to the moving Voronoi mesh is its ability to capture fluid instabilities at relatively low resolution \citep{2011arXiv1109.2218S}.  It has also been suggested that the Voronoi mesh would be ideal for capturing turbulent flow, as the scale-by-scale structure of eddies within eddies seems naturally suited to a Lagrangian treatment \citep{2012MNRAS.423.2558B}.

However, this method is not without its detractors.  A major concern regarding the Voronoi method is the noise introduced at the grid scale.  First, the extremely well-preserved contact discontinuities can produce jagged edges at the grid scale, which might potentially introduce numerical issues similar to those produced by contact steepeners, which were first introduced to artificially enhance contact discontinuities \citep{1984JCoPh..54..174C}.

A much larger source of noise, however, comes from the mesh motion itself.  This has been directly observed in simulations of driven turbulence.  When the mesh is moved, a large amount of power is introduced artificially at the grid scale, hampering the code's ability to resolve the inertial range of turbulence at a given resolution \citep{2012MNRAS.423.2558B, 2014MNRAS.437..397M}.

The cause of this noise is easily understood.  It is produced whenever two neighboring mesh points are given significantly different velocities.  In this case, mesh topology can change very abruptly, either producing diffusion when a face flips and overtakes a substantial fraction of a zone, or producing noise when a face rotates quickly about its center, moving fluid around artificially even when there is no flux through the face.

Such noise will always plague Voronoi codes like TESS and AREPO, as long as there is shear in the mesh at the smallest scales.  In this work, we report on a very simple way to reduce this noise, while still retaining all the advantages of the moving mesh.  Simply, rather than moving the mesh-generating points with the local fluid velocity $\vec v(\vec x)$, this velocity field is smoothed on some length scale, $\lambda$: $\vec w(\vec x) = S_{\lambda}\left[ \vec v( \vec x ) \right]$, and the mesh points are moved with the velocity $\vec w$, rather than the local fluid velocity.  In the examples presented here, the length scale $\lambda$ is of order the grid scale.  Note that in general, one always has complete freedom when choosing the mesh velocity field $\vec w$; if $\vec w = 0$ everywhere, the method reduces to an Eulerian fixed-grid scheme.  By smoothing $\vec w$, the code remains Lagrangian on large scales, but behaves like a comoving Eulerian code on small scales.  The idea of using a smoothed velocity field to move mesh points has previously been proposed in the context of SPH, in the form of a technique called X-SPH \citep{1989JGR....94.6449M, 1992ARA&A..30..543M}.

It is worth noting that both TESS and AREPO already employ mesh-regularization techniques with the goal of keeping the Voronoi tessellation regular (i.e. the Voronoi cells are ``rounder" and have fewer neighbors on average).  This work aims to treat a different problem with a distinct solution; the mesh velocity field is smoothed with the goal of reducing noise from rapid mesh adjustments.  In doing so, there is also the side-benefit that the mesh tends to remain more regular.  However, the reason for the smoothed mesh-motion is a reduction in small-scale noise.


This adaptation is simple, and it can provide improvement to the original method, as we shall demonstrate with several code tests.  It is recommended that every Voronoi code use a smoothed velocity field for the mesh motion, otherwise such codes are succeptible to the aforementioned numerical noise, and the code's ability to accurately capture instabilities and turbulence (which can have shear on all scales) may be in question.

\begin{figure}
  \centering
  \includegraphics[width=0.8\columnwidth]{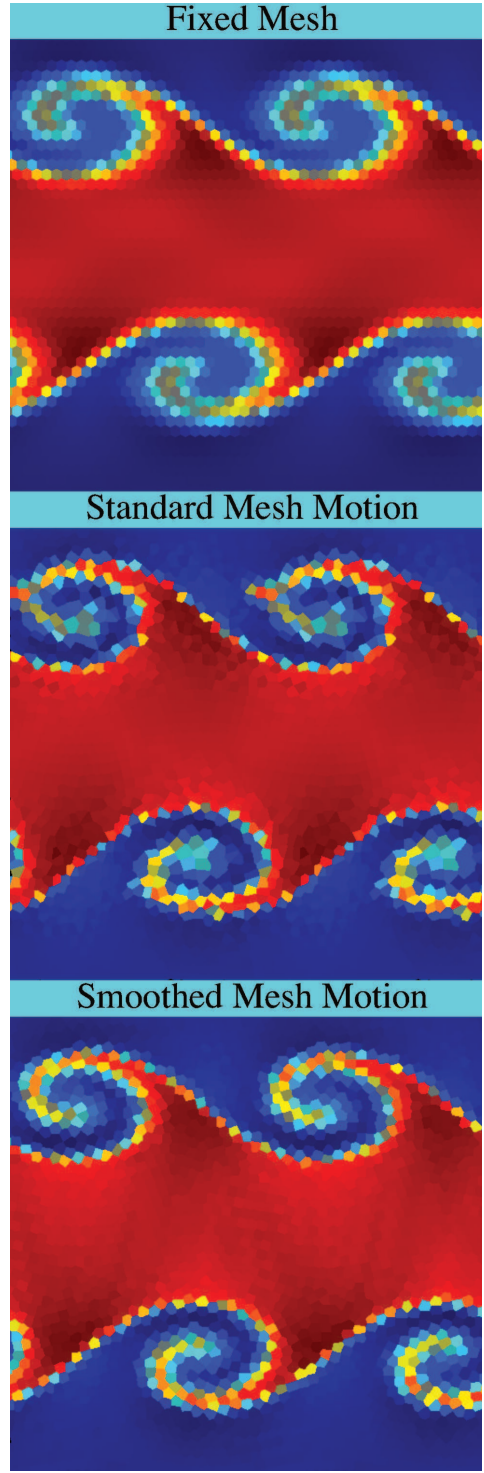}
  \caption{ Kelvin Helmholtz Instability, identical to the test performed in 
Springel (2010).  This snapshot is at time $t=1.5$. The upper panel uses a fixed mesh, which diffuses the eddies.  In the center panel, the mesh is moving, and the spiral shapes formed are sharper and more pronounced.  However, significant noise is introduced at the grid scale which is enhanced by the unstable shear motion, producing jagged structures.  The lower panel uses the smoothed mesh motion, which significantly reduces this noise, while still maintaining contact discontinuities and capturing the growth of the instability to high accuracy (See Figures \ref{fig:kh} and \ref{fig:growth}).}
  \label{fig:50}
\end{figure}

\section{Smoothed Mesh Motion}
\label{sec:smooth}

The moving Voronoi mesh technique has been articulated in other works \citep{2010MNRAS.401..791S, 2011ApJS..197...15D, 2011arXiv1109.2218S}, and we shall not repeat all of the details here.  Essentially, it is a finite volume method in which the finite volumes are calculated from a Voronoi tessellation of a given set of mesh-generating marker points.  These marker points can be given an arbitrary mesh velocity $\vec w(\vec x)$, though typically their motion is set to the local fluid velocity, $\vec w(\vec x) = \vec v(\vec x)$.  Motion of the marker points results in mesh motion, which is accounted for by adjusting the flux through each Voronoi face, based on the face's velocity.

Note that the velocity of a face does not have to mimic the velocity of nearby marker points.  In particular, if two neighboring marker points have significantly different velocities, this can result in a relative face velocity in a totally different direction from the velocity of the marker points.  This generally results in a face which moves counter to the flow, which is not desirable.  It can result in diffusive mixing if the face overtakes a significant fraction of one of the zones during a timestep.  It can also result in noise if the face rotates about its center considerably over a timestep, an action which is not accounted for by any compensating fluxes.

This problem can be mitigated by reducing the relative velocity of neighboring marker points.  In other words, instead of setting $\vec w$ equal to the local velocity, this velocity field is first processed through some smoothing operation, $\vec w = S_{\lambda}\left[ \vec v \right]$, so that neighboring marker points have similar velocities, while large-scale fluid motion is still properly subtracted off by the smoothed velocity field of the mesh.

In principle, there is freedom in the choice of the operator $S_{\lambda}$.  In this work, a very simple operation is employed.  Other smoothing operators are certainly possible; the choice of $S_{\lambda}$ presented here is mainly a proof-of-concept.  First, the marker point velocities are set equal to the local fluid velocity, $\vec w = \vec v$.  Next, an averaging operation is performed: $\vec w \rightarrow \alpha \vec w + (1-\alpha) \left< \vec w_j \right>$, where $\left< \vec w_j \right>$ represents a weighted average of the velocities of neighboring marker points (weighted by the area of the face shared in the Voronoi tessellation), and $\alpha$ is an adjustable constant chosen based on how aggressively the mesh is to be smoothed.  It is then possible to perform this substitution iteratively until the velocity field $\vec w$ has the desired smoothness.  In the examples presented in this work, $\alpha = 0.5$ is chosen with five iterations, but other smoothing operators could be chosen which may have more desirable properties.

\section{Code Tests}
\subsection{The Kelvin-Helmholtz Instability}
\label{sec:kh}

\begin{figure}
\centering
\includegraphics[width=0.8\columnwidth]{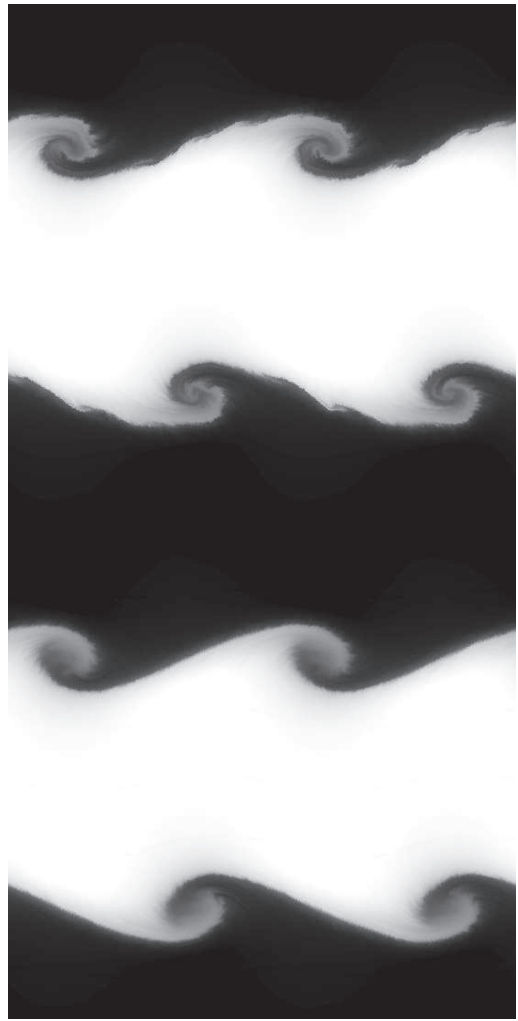}
\caption{ Smooth 2D Kelvin-Helmholtz test; initial conditions found in McNally et al. (2012) (snapshot at $t=1.5$, to be compared with Figure 2 of McNally et al. (2012)).  Both panels ran at a resolution of $512^2$.  The upper panel uses the standard mesh motion, while the lower panel uses smoothed mesh motion.  Naively one might incorrectly interpret the upper panel as being more accurate; the image ``looks" more realistic, as it appears to capture details that the lower panel misses.  However, since this is a smooth problem, there is a highly accurate reference solution found in McNally et al. (2012), calculated using the high-order Pencil code at high resolution.  Their result does not exhibit the small-scale features found in the upper panel.  Rather, these are secondary instabilities, which are artificially seeded by mesh noise.}
\label{fig:kh} 
\end{figure}

\begin{figure}
\centering
\includegraphics[width=1.05\columnwidth]{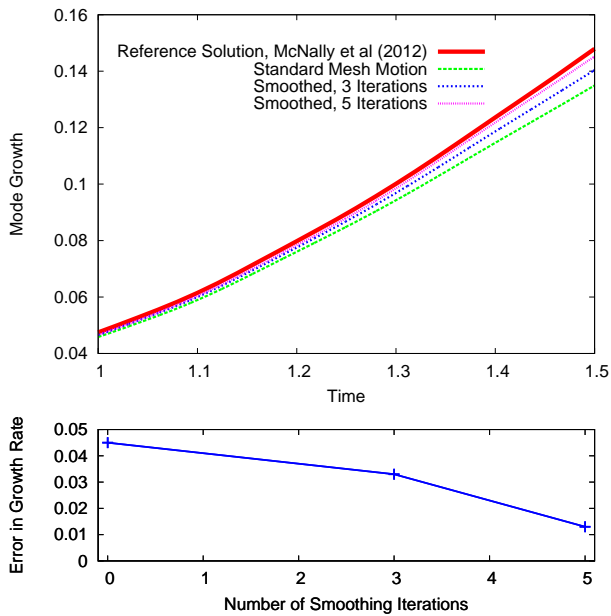}
\caption{ Mode growth of the smooth Kelvin-Helmholtz test of Figure \ref{fig:kh} (definitions and reference solution found in ).  Calculations used a mesh with $512 \times 512$ zones, and the effect of smoothing is tested by varying the number of iterations in the smoothing operator.  Grid noise suppresses the growth of the instability, but by smoothing the mesh motion the growth rate is more accurately captured.}
\label{fig:growth} 
\end{figure}

\begin{figure*}
\centering
\includegraphics[width=1.6\columnwidth]{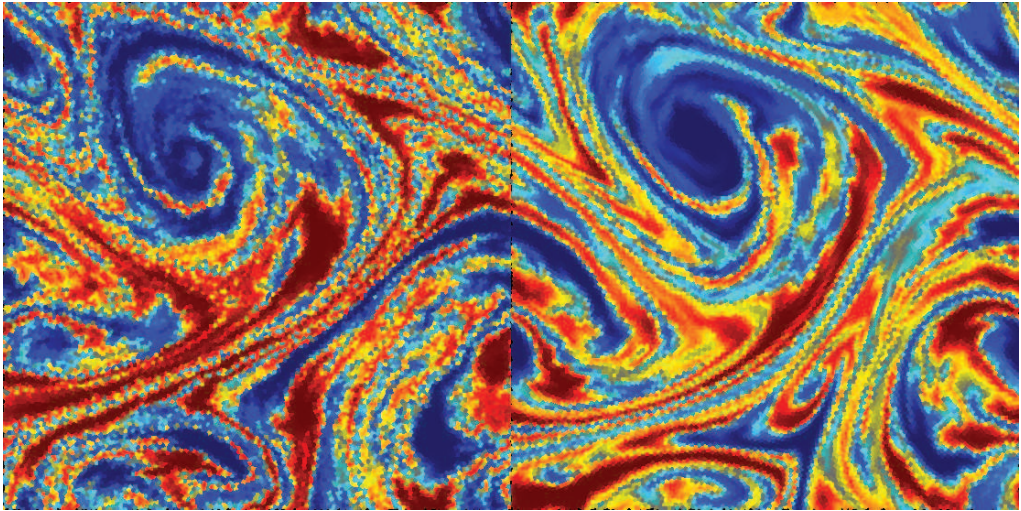}
\caption{ Simulations of driven 2D turbulence produced by a stochastic driving field.  The fluid motion is made visible by the means of a passive scalar field which was initially given by a step function in the x coordinate.  The two panels are identical, except that the left panel employed standard mesh motion, while the right panel used the smoothed-mesh prescription.}
\label{fig:turb} 
\end{figure*}

The Kelvin-Helmholtz instability was at first cited as one of the best arguments for using a moving Voronoi mesh.  \cite{2010MNRAS.401..791S} showed that the eddies formed in the instability are much more vividly captured when moving the mesh than when the mesh is kept fixed, since a fixed mesh introduces advection errors which diffuse the eddies.  Figure \ref{fig:50} shows a very low-resolution ($50 \times 50$) Kelvin-Helmholtz calculation at $t=1.5$ based on the initial conditions given in \cite{2010MNRAS.401..791S} (Springel's Figure 32).  The same calculation is performed on a fixed mesh (top panel), standard moving mesh with $\vec w = \vec v$ (center), and finally smoothed mesh motion with $\vec w = S_{\lambda}[\vec v]$ (lower panel).  It is clear by looking at the top and center panels that moving the mesh results in a dramatic reduction in diffusion, and much sharper contact discontinuities.  The fixed mesh exhibits diffusion which comes from advection errors to which the moving mesh is less susceptible.

However, upon closer inspection, the standard moving mesh exhibits a different problem.  Small-scale structures appear to form in the eddies, which may be evidence for spurious noise or numerical instability (similar features can be seen in \cite{2010MNRAS.401..791S}, Figure 32).  At first glance, it might seem as though these small structures could actually be real; i.e. that they might be secondary instabilities that moving-mesh codes can capture, but which are diffused away by fixed-grid codes.  In this section we will demonstrate that this is not the case.  That is, secondary instabilities do exist, but in these examples they are seeded artificially by the grid-scale noise described in the previous section.  The unstable nature of the flow enhances the noise and makes convergence slow.  However, when the mesh motion is smoothed (lower panel of Figure \ref{fig:50}), the noise is reduced and the result is more accurate.  Because the smoothing operation is not significantly diffusive, one does not have to sacrifice the sharpness of the contact discontinuities.

The calculation with smoothed mesh motion can be shown to be correct, using the smooth Kelvin-Helmholtz test proposed by \cite{2012ApJS..201...18M}.  These authors started with a smooth initial condition, and evolved it using the 6th-order Pencil code \citep{2010ascl.soft10060B} at a resolution of $4096^2$.  This resolution study was careful and thorough, and therefore this is likely the closest one will come to having an exact solution to this problem (relative errors in their results were of the order $10^{-3}$ or smaller everywhere on the grid; far too small to see by eye).

Figure \ref{fig:kh} shows a snapshot of the TESS code performing this problem at a resolution of $512^2$.  The upper panel uses standard mesh motion, and the lower panel employes smoothed mesh motion.  To the careless eye, the upper panel ``looks" more correct, as it seems as if it is picking out more details than the lower panel.  However, the true, converged solution (Figure 2 in \cite{2012ApJS..201...18M}) does not exhibit these detailed features.  These are secondary instabilities seeded by numerical noise.  Moreover, it is very difficult to eliminate these errors with high resolution; increasing the resolution introduces the noise on smaller scales, which excites faster-growing modes.  The amplitude of these errors eventually trends to zero with high resolution, but slowly.

Fortunately, smoothing the velocity field of the mesh significantly reduces this problem.  Figure \ref{fig:growth} shows that the growth of the instability is more accurately captured when using smoothed mesh motion.

\subsection{Driven Turbulence} 
\label{sec:turb}

\begin{figure}
\centering
\includegraphics[width=0.995\columnwidth]{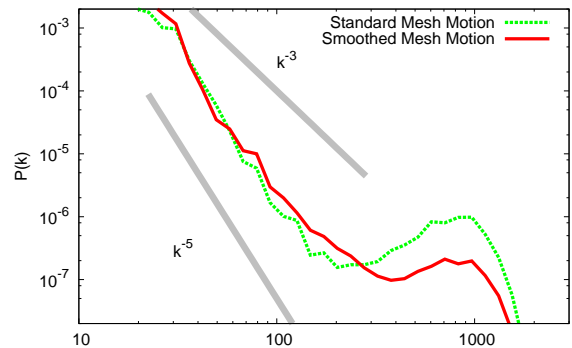}
\caption{ Power spectra of driven 2D turbulence using a mesh of $512 \times 512$ zones.  There is a clear inertial range extending over roughly an order of magnitude in k.  The inertial range shows a steep scaling somewhere between $k^{-3}$ and $k^{-5}$.  At small scales, there is significant power at the dissipation scale, indicating noise induced by mesh motion.  This noise is reduced and the inertial range is extended when using a smoothed velocity field for the mesh.}
\label{fig:power} 
\end{figure}

\begin{figure}
\centering
\includegraphics[width=0.99\columnwidth]{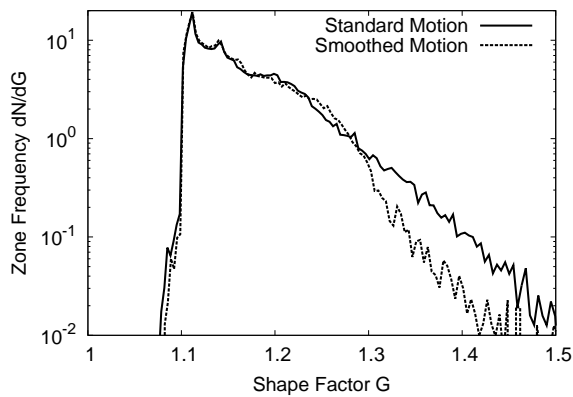}
\caption{ Statistics of zone shapes, showing the relative frequency of zones with a given shape factor $G$ (proportional to perimiter squared over area).  Zone data is taken from a $256^2$ run of the smooth KH test at $t=1.5$.  $G=1$ describes a circle, while $G=1.1$ for a regular hexagon.  Smoothed mesh motion reduces the number of zones in the tail, around $G = 1.4$, by nearly an order of magnitude.}
\label{fig:shape} 
\end{figure}

The smoothed mesh motion is important in any problem which exhibits shear on all scales.  In a turbulent cascade, there is always shear on all scales, and therefore substantial grid noise will be produced when using a Voronoi mesh to follow such a flow.  If the mesh is moved, but with a smoothed prescription for the mesh motion, the mesh follows the largest eddies, and this large-scale flow is effectively subtracted off.  Meanwhile, the dynamics of the smallest eddies are calculated in an effectively Eulerian manner (in the smallest eddy's center-of-momentum frame).  Thus, the turbulence is captured in a natural way, but without introducing the noise produced by shearing the grid on the smallest scales.

This can be illustrated in two-dimensional simulations of driven turbulence.  The fluid is stirred on large scales using a driving field generated via a stochastic process.  This method for large-scale driving is identical to the one employed by \cite{2012ApJ...744...32Z}.  The resultant turbulence is subsonic (Mach $0.1$).  A snapshot of the 2D turbulence is shown in Figure \ref{fig:turb}, where a passive scalar is included to reveal the structures generated from the turbulent fluid motions.  This passive scalar was initialized as a step function.  The advantage gained in smoothing the mesh motion can be seen most clearly in the power spectrum (Figure \ref{fig:power}).

In the power spectrum, there is a clear inertial range, extending from the driving scale at $k \sim 10$ to a bottleneck at small scales.  In 2D, on scales smaller than the driving scale, the power spectrum is set by an enstrophy cascade from large to small scales.  Simple scaling arguments give a scale dependence of $k^{-3}$ for the power spectrum in the inertial range \citep{1967PhFl...10.1417K}, but it has been suggested that the nonlocality of 2D turbulence may cause this slope to depend on the fluid Reynolds number \citep{2010JFM...646..517B}.  The important question is not of slope of the spectrum but on the extent of the inertial range and the size of the bottleneck at the dissipation scale.  Clearly, the smoothed mesh motion reduces the power in this bottleneck, and increases the extent of the resolved inertial range.  This could potentially be further improved with a different choice of $S_{\lambda}$.

\section{Summary}
\label{sec:sum}

The noise introduced by the Voronoi mesh motion at small scales can hamper the convergence properties of codes which employ this moving-mesh technique.  Calculations which are sensitive to the detailed development of instabilities and turbulence should employ some mechanism for reducing or eliminating this noise, such as the smoothed mesh motion described in this work.

More generally, when performing astrophysical or cosmological calculations using a moving Voronoi mesh, if a certain qualitative behavior depends on whether the mesh motion is turned on, it is not necessarily the case that the moving-mesh version gives the correct answer.  In the case of the smooth Kelvin-Helmholtz example of Figure \ref{fig:kh}, it may be very tempting to assume that these small structures are subtle details being picked out by the mesh motion.  However, the resolution study by \cite{2012ApJS..201...18M} clearly demonstrated that this was not the case.  This strongly suggests that the same is true of the non-smooth version shown in Figure \ref{fig:50}.

Given this, the smoothed mesh motion presented here provides a more accurate solution for the Kelvin-Helmholtz problem, without sacrificing the advantages gained by moving the mesh; this improvement came from subtracting noise, not from adding diffusion.  The turbulence calculations (Figures \ref{fig:turb} and \ref{fig:power}) provide more evidence for this; the mesh noise is clearly exhibited in the high-k end of the power spectrum.  This loud bottleneck at small scales is reduced by smoothing the mesh motion.

It should also be noted that while mesh smoothing is a distinct operation from the ``mesh steering" techniques used in AREPO and TESS (including the mesh-steering method of \cite{2012MNRAS.425.3024V}), it can help to accomplish similar goals.  Figure \ref{fig:shape} shows the distribution of zone shapes, defining the ``shape factor"

\begin{equation}
G = {\rm Perimiter^2 \over 4 \pi (\rm Area)}.
\end{equation}

$G$ represents how far off zones are from being perfectly round ($G = 1$ in the limiting case of a circle, and $G = 1.1$ for a regular hexagon, but $G$ can potentially attain much larger values for non-round shapes and large aspect ratios).  The figure shows that with standard mesh motion, mesh steering already does a reasonable job keeping zones round.  However, smoothed mesh motion improves on this a bit, reducing the number of zones with $G \sim 1.4$ by nearly an order of magnitude.

Finally, it should be noted that the smoothing operator $S_\lambda$ presented in section \ref{sec:smooth} is not unique; undoubtedly there are other operations one can perform to smooth the velocity field $\vec w$ on small scales, which may improve performance further.

This research was supported in part by NASA through grant NNX10AF62G issued through the Astrophysics Theory Program and Fermi grant NNX13A093G and by the NSF through grant AST-1009863.  We are grateful to Jim Stone, Colin McNally and Jonathan Zrake for helpful comments and discussions.

\bibliographystyle{mn2e}

\end{document}